# Requirements Rationalization and Synthesis enabled by Model Synchronization

Siyuan Ji, *Member, IEEE*, Charles E. Dickerson, *Senior Member, IEEE*, Michael Wilkinson

*Abstract*— In the international standard for system and software engineering ISO/IEC/IEEE 15288: 2015, the output of the stakeholder needs and the business or mission analysis technical processes are transformed into a technical view of the system by the system requirements definition process. In model-based systems engineering, functional needs can be modeled by use case diagrams. Intended outcomes of system requirements definition include resolution of disagreement about requirements, explicit agreement between stakeholders, and traceability. However, stakeholder needs are often elicited in a siloed manner and may be inconsistent. The lack of mathematically based systematic approaches for requirements definition poses a challenge to model-based transformation of needs into a technical view of the system that achieves agreement between stakeholders. This paper specifies and demonstrates mathematical frameworks for *rationalizing* and *synthesizing* functional needs that have been captured through an elicitation process. Benefits of this approach include but are not limited to supporting rigorous identification and resolution of disagreements and facilitating systematic analysis of change impact to achieve stakeholder agreement all with minimal intervention by the system engineers.

## I. INTRODUCTION

SYSTEM requirements definition processes such as specified in ISO/IEC/IEEE 15288: 2015 [1] seek explicit agreement between stakeholders on system requirements and their traceability. Terminology such as rationalization, synthesis and synchronization of models are not used in the standard but are frequently used by the community to express fundamental concepts and processes for systems and software engineering. See for example the usage of synthesis in the INCOSE SE Handbook pp. 11, 157, 201 [2]. The Object Management Group (OMG) prescribes standards for software technologies such as the Systems Modeling Language (SysML) [3] but similar to other standards bodies, they do not prescribe methods for implementing their standards. Recent research by the authors has investigated methods for agreement and traceability processes in systems engineering using model synchronization [4].

### A. Background and Motivation

Stakeholder needs are often elicited in a 'siloed' manner, in which systems engineers identify concerns and elicit needs from different groups of stakeholders in isolation. This practice supports definition of architecture viewpoints that address the different stakeholder concerns [5]. Stakeholder requirements derived from different viewpoints must be analyzed for completeness and consistency, among other qualities. Issues such as conflicting needs must be resolved. The stakeholder views must be agreed upon as part of requirements management [1]. In practice, this involves identifying and resolving issues such as conflicting needs from different stakeholders. It can be enabled through stakeholder workshops in which the outcome achieved is an agreement among stakeholders that documents a final synthesis of their needs and requirements. The outputs of the technical processes for stakeholder needs and the business or mission analysis must be of sufficient engineering quality to be used as the basis for system requirements specification.

It is widely acknowledged that errors made during the early stages of the lifecycle, say concerning needs or architectural choices, may have significant impact in later stages, see for example [6-8]. Well-known general model-based techniques and methodologies [9-11] are seen to provide a good starting point for identifying and resolving incompatible needs. Graphical models, e.g., SysML [3], are also intended to facilitate stakeholder communications to achieve agreement. However, these methods do not offer rigorous methods to support a managed agreement process of stakeholder inputs to system requirements definition. If MBSE in general is to be adopted effectively in the practice of stakeholder requirements elicitation and management, the diagrams that support stakeholder communication, e.g., created in SysML, must be supported by rigorous methods. This is a key motivation for addressing requirements in a rigorous and repeatable manner using a mathematical framework. Use of such a framework could also help with issues known to arise in later stages of the lifecycle [12]. Such an approach could also help to address from an early stage what other authors have identified as emergent complexity generating properties in engineering design [13,14].

(*Corresponding Author*) Siyuan Ji is with the Wolfson School of Mechanical, Electrical, and Manufacturing Engineering, Loughborough University, Epinal Way, Loughborough, UK (e-mail: s.ji@lboro.ac.uk).

Charles E. Dickerson is with the Wolfson School of Mechanical, Electrical, and Manufacturing Engineering, Loughborough University, Epinal Way, Loughborough, UK (e-mail: c.dickerson@lboro.ac.uk).

Michael Wilkinson is with BAE Systems Submarines and a visiting professor at the Wolfson School of Mechanical, Electrical, and Manufacturing Engineering, Loughborough University Epinal Way, Loughborough, UK (e-mail: m.k.wilkinson@lboro.ac.uk).

Color versions of one or more of the figures in this article are available online at http://ieeexplore.ieee.org

Agreement can be a resource-intensive process. With each of many stakeholders advocating for their own needs, it can be difficult to achieve agreement. MBSE seeks to make the process more efficient and effective by using models as the focal point of communication and analysis in order to facilitate agreement. Even so, there are interrelational dependencies between needs that can lead to possibly endless cycles of resolution and agreement.

Recent model theoretic advances [4] by the authors have demonstrated that structure preserving transformations of models developed using relational orientation can be synchronized to accommodate changes made to models not just in progress but also at a later time. We further explore these advances by using relational methods to enable model-based agreement between stakeholders. This is a promising approach to support efficient and effective stakeholder meetings for the identification and resolution of requirements issues, and synthesis of the stakeholder needs.

It should be noted that this paper does not attempt to define a strictly formal approach to requirements analysis (as advocated for example in [15-17]); rather, it uses a mathematical framework to assist the analyst within a joint cognitive paradigm [4]. Within the context of the broader research programme of the authors [18], we believe the joint cognitive paradigm begins to address in a pragmatic way calls for greater rigor in the development and application of Systems Engineering [19-22].

*B. Definition and Discussion of Terminology*

To facilitate the description and understanding of the framework and process proposed in this work, we define the following key terms based on common usage (derived from dictionary definitions and technical usage) and intended usage in practical engineering, in a model-based context:

*Synchronization* – A process that establishes and maintains the consistency of the elements and semantics occurring between two models. A pair of models is synchronized when the models are free of issues such as unnecessary repetitions or contradictions.

The common usage of the term synchronize refers to entities that occur or operate at the same time. In software engineering, the term is focused on multiple programmers working on components of a body of computer code. In Ref. [4], this concept was elevated to a higher level of abstraction for systems.

*Rationalization* – A process that seeks to find logical reasons to bring a set of models into agreement.

There are many concepts and definitions of this term. *Harmonization* is a similar term. In general software engineering, the phrase *application rationalization* is defined as a process of reviewing an application inventory to determine which ones should be retired, retained, reposted, replatformed, refactored or reimagined [23].

As used in this paper, Rationalization is a process that brings a set of models to a state that is free of incompatible model elements. Incompatibility in the context of requirements engineering and MBSE refers to the following two cases: (i) structurally incompatible e.g., the hierarchical relationship between two requirements is inconsistently modelled in two places; or (ii) contradictory or contrary e.g., where two requirements as specified in two different models cannot be simultaneously true. A more technical definition of Rationalization based on logic and set theory will be proposed in Section II-A.

In this paper, we will only consider *endogenous* rationalization (adapting from the terminology of endogenous transformation [24] used in model-driven software engineering) of graphical models that are of the same type, e.g., Use Case diagrams (UCDs). Models in a rationalized state attained through this process can be synthesized in a seamless manner, which in this paper is achieved by a straightforward model merging. This leads us to the third term.

*Synthesis* – A process that composes parts (partial models) to form a whole (a complete model). Synthesis defined in this way has been used in various other contexts, such as the synthesis of musical notes into a piece of music. In requirements engineering, this can be understood as the idea of combining different stakeholder views and their needs into a consistent 'system of stakeholder requirements', captured in a single model.

*C. Key Contributions of This Work*

We claim two key contributions brought by this work:
1) **Framework for Rationalization and Synthesis:** Model theoretic foundations that facilitate rationalization and synthesis of stakeholder needs.
2) **Rationalization Process enabled by Synchronization:** A matrix-based process that guides the rationalization of stakeholder needs, which enables the transformation of the outcomes of an elicitation process into a set of *rationalized* stakeholder requirements models.

To demonstrate how the framework and process works in practice and the benefits they bring, we will apply the process in a case study of a robotic arm, in which siloed stakeholder needs are rationalized and synthesized initially, and further re-rationalized following changes having been introduced to the models.

*D. Structure of the Paper*

The rest of this paper is structured as follows: in Section II, we introduce the theoretical framework that underpins model rationalization and model synthesis as defined in this section, and construct rationalization and synthesis processes that can be adopted and executed in practical settings. Practicality is then illustrated through an incremental application of the processes to a robotic arm case study in Section III. To further demonstrate the synchronization property of the rationalized models, Section IV continues the case study by introducing requirements changes and demonstrates how the changes can be analyzed and accepted efficiently while maintaining consistency between the models. Finally, in Section V we discuss the practicality of our research for various model-based methods of requirements elicitation. The Conclusion provides a critical analysis of the results and summarizes future work.

## II. Foundations and Methods

This section introduces a foundation for the proposed rationalization, synthesis, and synchronization processes in this paper. We employ the mathematically based ROSETTA Framework developed by Dickerson and Mavris [25] and first applied in [26]. Our method is model theoretic as per Tarski and Predicate Calculus. Representing the models and their transformation in matrix representations is a mathematical technique that is precise and provides a basis for rigorous development of algorithms.

We assume that following a typical model-based approach for requirements elicitation (see Section V for related works), the outcomes are graphical models constructed using standardized modelling language such as the SysML. In such graphical models, model elements are either entities or relations, allowing one to treat them as simple graphs of vertices and edges, and thus represent them using matrices, with matrix elements showing relations between entities. Furthermore, it is possible that the elicitation process is done in situations where stakeholder needs are elicited in a siloed manner. As such, stakeholders are not necessarily aware of the needs articulated by other groups. This results in the need for processes to identify and resolve incompatible requirements, and to produce a 'single source of truth' capturing the agreed sets of stakeholder needs for the derivation of system requirements.

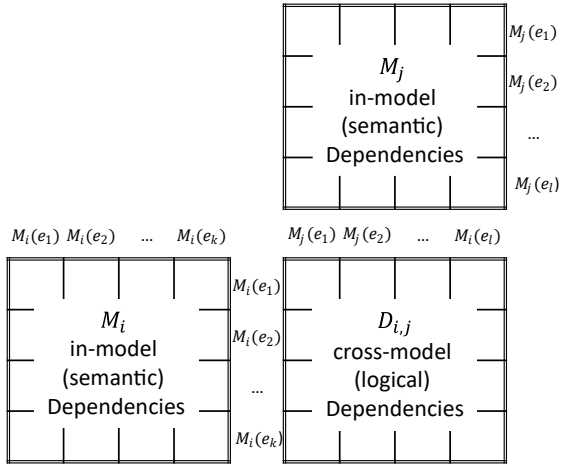

**Fig. 1.** Framework for rationalization.

### A. The Framework for Rationalization

Based on the ROSETTA Framework, the proposed framework for rationalization, denoted by a $M_i$-$D_{i,j}$-$M_j$ triad, is constructed and depicted in Fig. 1. This framework uses two adjacency matrices, $M_i$ (bottom-left) and $M_j$ (top-right) to represent respectively the elements in two graphical models of the same type. We use the convention, $M_i(e_x)$ to represent an entity in the corresponding model, while the matrix elements in $M_i$ capture the relations between two such entities; these relations are referred to as in-model dependencies. The semantics of the matrix elements depends on the metamodel governing the graphical model. Taking UCDs as an example, in-model dependencies include Association (between an Actor entity and a Use Case entity), Allocation (from a Use Case to the System), Inclusion, Extension and Usage (which are all between two Use Cases).

The framework further utilizes an incidence matrix, denoted as $D_{i,j}$ (bottom-right), to represent the so-called cross-model dependencies [4] between the entities of the models, i.e., relations between $M_i(e_x)$ in $M_i$ and $M_j(e_y)$ in $M_j$. These dependencies are logical in nature. Thus, based on the formal logic of possible (logical) dependencies between two elements, the following types of relations can be formulated between two model elements from different models:

- **E** – Equivalence: $e_x = e_y$
- **H** – Hierarchical: $e_x \subset e_y$
- **O** – Overlapping: $\exists e_z, e_z \subset e_x, e_z \subset e_y$
- **I** – Implication: $e_x \rightarrow e_y$
- **C** – Contradiction: $\neg(e_x \rightarrow e_x)$
- Unrelated: no logical dependencies

This is not an exhaustive list of logical dependencies, but it is sufficient for the applications in this paper and indeed for most engineering problems. For instance, **H**-, **O**-, and **I**-dependencies indicate potential structural incompatibility between the two model elements, where there is a relationship between model elements across the models not being captured by either of the models. On the other hand, incompatible requirements that are conflicting in nature can be identified as having a **C**-dependency. Of the above dependencies, **H**- and **I**-dependencies are directional. Directionality is captured using a directed arrowhead in the subscript, e.g., **I**$_\rightarrow$ for $M_i(e_x) \rightarrow M_j(e_y)$ and **H**$_\leftarrow$ for $M_i(e_x) \supset M_j(e_y)$.

Our technical definition of a *rationalized state* between two models is given below: Two models are rationalized when the $D_{i,j}$ matrix contains only **E**- (equivalence) dependency or void (empty) matrix elements, representing equivalent and unrelated entities, respectively.

### B. The Framework for Synthesis

Using matrix representation, the framework for synthesis is constructed and depicted in Fig. 2. This framework, unlike the previous, is a matrix of matrices, where the header of a row or a column now represents a set of entities contained in a model, $M_i$. For instance, the notation $M_i(e)$ describes a row (column) vector when it is used as the header of a column (row). As such, every block (element) in this matrix, is representing a matrix, either denoted by $M_i$ for the diagonal blocks or $D_{i,j}$ for the off-diagonal blocks. The notations are self-explanatory referring back to the framework for rationalization as depicted in Fig. 1. Notably, the matrix of matrices is symmetric. This is because dependencies, whether in-model or cross-model, are directional such that $D_{i,j}$ and $D_{j,i}$ have exactly the same semantic content. Therefore, the framework is simplified down to an upper triangular matrix of matrices as shown in Fig. 2.

For a set of *n*-models, *Synthesis*, as a process, is technically understood as merging the models. In the case where models are not rationalized, one would anticipate seeing various logical dependencies in the $D_{i,j}$ matrices. These elements would make merging difficult and in certain cases, infeasible. For example, where there is a contradiction, **C**-dependency, merging is impossible since the presence of one model entity contradicts the presence of the other. Arguably, one could rationalize the models during synthesis. However, in this approach, a resolution of problematic logical dependencies, such as a **C**-dependency, could generate new **C**-dependencies elsewhere in another $D$ matrix, making it difficult to track changes and re-rationalize.

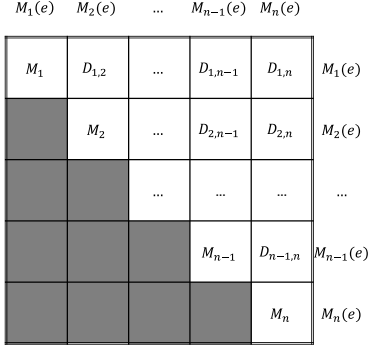

**Fig. 2.** Framework for synthesis.

A better alternative is to rationalize the models using model synchronization prior to synthesis. When models are all synchronized with each other, according to the framework for rationalization, $D_{i,j}$ matrices will contain only **E**-dependencies as non-empty matrix elements. Then, synthesizing the rationalized models becomes seamless because the equivalent cross-model entities will merge into a single entity. In terms of matrix operation, an **E**-dependency suggests that a row/column is repeated, thus allowing repeated row/columns to be merged into one, inheriting all the in-model dependencies prior to the merge. Such merging can be readily implemented by existing technologies in software engineering [27].

*C. The Rationalization & Synthesis Processes*

In principle, rationalization and synthesis can be treated as two independent processes, e.g., it is possible to synthesize models without rationalizing them first. However, following the frameworks introduced in the previous section, it is evident that a synthesis preceded by an endogenous rationalization of models of the same type is advantageous; hence we construct the two processes in series, as conceptually illustrated in Fig. 3, where we have adopted the following notations:

- $M_i^0$    Initial model that are not necessarily rationalized
- $M_i'$    Rationalized model
- $M$    Synthesized model
- $D_{i,j}^0$    Initialized cross-model dependencies; empty
- $D_{i,j}$    Cross-model dependencies pre-synchronization
- $D_{i,j}'$    Cross-model dependencies post-synchronization
- $\mathcal{J}$    Process: identification of cross-model dependencies
- $\mathcal{S}$    Process: synchronization
- $\mathcal{M}$    Process: merging
- $\rightarrow$    Process flow
- $\dashrightarrow$    Linkage for a synchronized triad

Note that (i) technically, $M$ and $D$ are matrices, with $M$ representing a corresponding model; (ii) $D_{i,j}^0$ is always initialized as an empty matrix; this does not mean the models are rationalized, hence no dashed arrows are drawn between $D_{i,j}^0$ and $M$s.

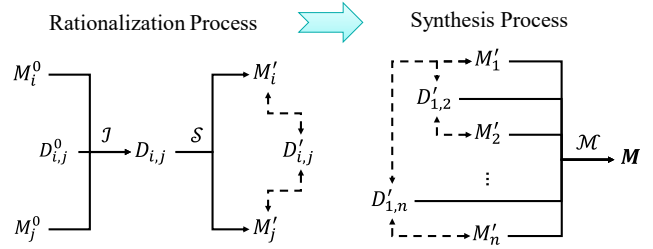

**Fig. 3.** A rationalization and synthesis process enabled by model synchronization.

The rationalization process starts with initializing $M_i^0$-$D_{i,j}^0$-$M_j^0$, followed by the identification of cross-model dependencies according to the prompt list of logical dependencies provided in Section II-A. These dependencies are captured in an intermediate matrix, $D_{i,j}$, which will serve as the basis for synchronization. The goal of the synchronization subprocess, as discussed in the framework, is to revise the models until $D_{i,j}$ is transitioned into a $D_{i,j}'$ where the $D_{i,j}'$ only contains **E**-dependencies as non-empty matrix elements. When this is achieved, a synchronized $M_i'$-$D_{i,j}'$-$M_j'$ triad is formed, marking the completion of the rationalization between the two (revised) models as respectively represented in $M_i'$ and $M_j'$. This is why the rationalization process is said to be enabled by model synchronization. Then, to complete the rationalization among all models, the rationalization process is repeated until all pairs of models are synchronized, as reflected through a dedicated triad.

Following the completion of rationalization among all models, the synthesis process merges all the triads together into an overarching model, $M$, that contains all the model elements. This is a seamless process as the merge is only concerned with merging repeated model entities (related by **E**-dependencies). Each merged entity inherits every in-model dependency that the entity has in different models prior to merging.

III. CASE STUDY PART I – RATIONALIZATION AND SYNTHESIS

This section demonstrates an incremental application of the rationalization and synthesis processes to a robotic arm case study.

*A. Functional Needs for a Robotic Arm*

In this subsection, we describe the settings and assumptions made for the robotic arm case study. The engineering project is about designing or acquiring a suitable robotic arm for a production line in a factory. To reduce the scope of the study, we will focus only on stakeholder functional needs, which can be modelled using UCDs in a typical model-based stakeholder needs elicitation approach.

During the elicitation, we assume that three stakeholders and their needs are identified initially:
- The Production Line Owner, who requires the robotic arm to be able to pick parts from one workbench and place them on another.
- Factory Planner, who, for the purpose of optimizing the factory floor and safety, requires the robot to be able to move its arm following a pre-defined path.
- Maintenance engineer, who needs to be able to test whatever functions the robot exhibits.

Assuming that the needs of individual stakeholders are elicited in a siloed manner, we derive three UCDs, each capturing the functional needs as Use Cases. The three UCDs are depicted in Fig.4, from left to right. It is worth noting that although the Production Line Owner and Factory Planner are modelled as Actors for the purpose of showing the siloed stakeholder view, they do not directly interact with the system; hence, no Associations have been used. We have also intentionally neglected the 'parts' being moved by the Robotic Arm as an Actor to simplify the models. In these diagrams, we have given each model entity a unique ID, denoted using the matrix convention introduced in Section II.

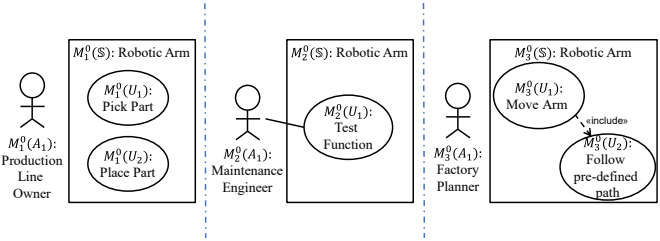

**Fig. 4.** A set of siloed UCDs for the Robotic Arm system; from left to right, UCD-1, UCD-2, and UCD-3, respectively.

*B. Rationalization of three models*

The proposed rationalization process needs to be applied to every possible pair of models, starting from the first pair, UCD-1 and UCD-2. The two models, together with yet to be identified cross-model dependencies, are initialized into a $M_1^0$-$D_{1,2}^0$-$M_2^0$ triad, produced and shown in Fig. 5, according to the framework for rationalization. The matrix representations use the ID conventions adopted in the diagrams. For $M_1^0$, we have the System, an Actor and two Use Cases, and the matrix elements capturing the semantic structure: allocations of the two Use Cases to the System. For $M_2^0$, we have the System, an Actor and a Use Case, and the matrix elements capturing the semantic structure: allocation of the Use Case to the System and association between the Actor and the Use Case. For $D_{1,2}^0$, this is empty to start with.

In the next step, we identify cross-model dependencies and update $D_{1,2}^0$ into a $D_{1,2}$ that captures these dependencies. Going through every pair of cross-model entities as guided by $D_{1,2}^0$, we have identified three dependencies, as shown on the left of Fig. 6. Firstly, there is the equivalence, **E**-dependency, for the System. This is straightforward as both diagrams are focusing on the same subject of interest: the 'Robotic Arm'. Secondly, an implication relation between the individual functionalities modeled in UCD-1 and the 'Test Function' functionality in UCD-2. Specifically, for every functionality, 'Test Function' implies that it needs to be tested, so the direction of implication goes from 'Test Function' to 'Pick Part' and 'Place Part', denoted by $\mathbf{I}_\leftarrow$ in the $D_{1,2}$ matrix to show the directionality.

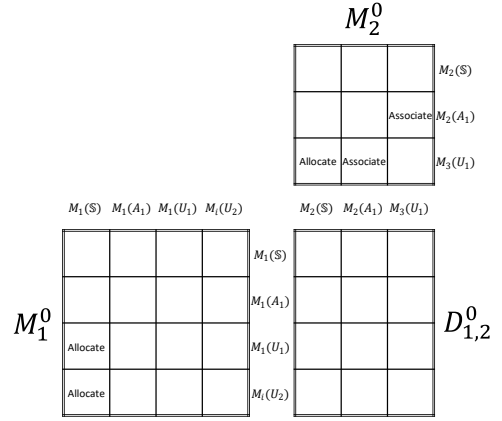

**Fig. 5.** Matrix representation prior to rationalization between $M_1^0$ and $M_2^0$; $D_{1,2}^0$ is empty prior to rationalization.

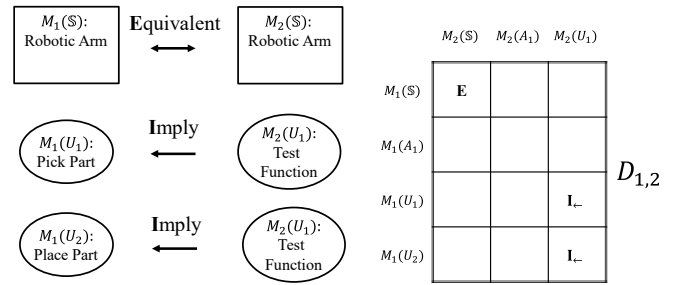

**Fig. 6.** Identified cross-model (logical) dependencies between entities in $M_1$ and entities in $M_2$, captured in $D_{1,2}$.

With $D_{1,2}$ derived, we then need to synchronize $M_1$ and $M_2$ until arriving at a $M_1'$-$D_{1,2}'$-$M_2'$ triad where $D_{1,2}'$ contains only **E**-dependencies as non-empty matrix elements. This synchronization sub-process can be intuitively understood as a resolution of 'disagreement' between stakeholders. Disagreements do not necessarily mean contradictions in this context but serve as points of interests that need to be investigated so 'agreement' can be achieved. Agreement is reflected through equivalent (repeated) functional needs appearing in different models, showing that different stakeholders agree on what they mean by that need.

In the case study, the repetition of Robotic Arm is expected, and thus, is already synchronized. For the two $\mathbf{I}_\leftarrow$-dependencies there could be different ways of resolution to achieve synchronization. For instance, one could make an argument that when the Maintenance Engineer tests a Robotic Arm's functionality, that functionality must be accessed (used); this can then be modelled by a Usage dependency with the «use» stereotype in an UCD. Another slightly more complicated, but more practical and meaningful resolution, is to introduce a normal mode and a test mode for each of the functions modeled in UCD-1. Instead of accessing the full functionality for testing, a dedicated test mode could make maintenance more efficient, but at the cost of higher complexity. In practice, how one achieves synchronization is subject to a trade-off analysis conducted during stakeholder discussions when different resolutions exist.

In this case study, we will be using the second resolution going forward. This resolution can be modelled by introducing a pair of included Use Cases for each of the primary functions, one dedicated to each of the two modes, as shown in the revised UCD-1 on the left of Fig. 7. To synchronize such changes made to UCD-1, the test model Use Cases are also modelled in the revised UCD-2, shown on the right of Fig. 7. The equivalent entities are self-evident.

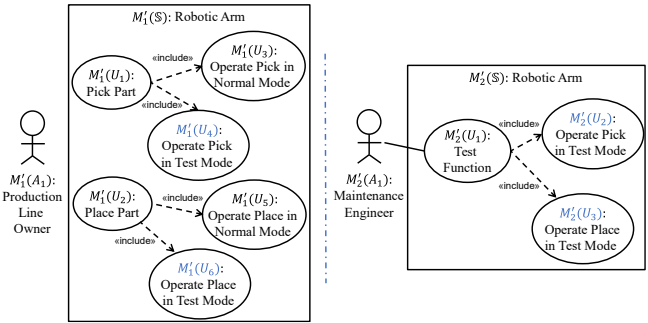

**Fig. 7.** Rationalized UCD-1 and UCD-2.

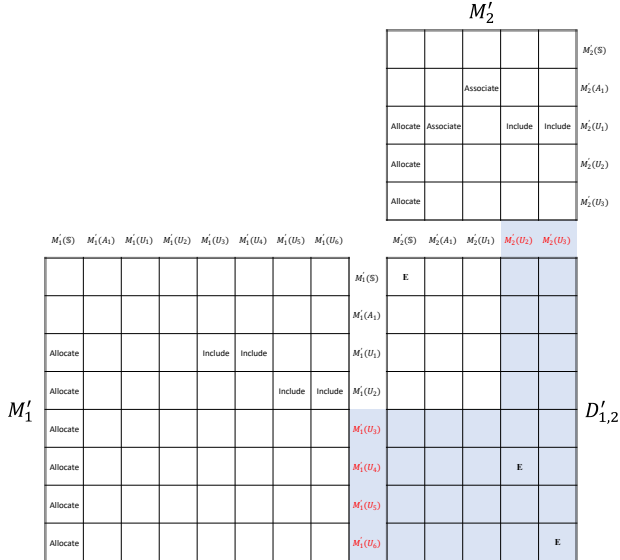

**Fig. 8.** The synchronized $M'_1$-$D'_{1,2}$-$M'_2$ triad for the rationalized UCD-1 and UCD-2; elements of $D'_{1,2}$ is either empty (unrelated) or **E** (equivalence).

Now that UCD-1 and UCD-2 are rationalized, the matrices can be updated to establish the synchronized triad, $M'_1$-$D'_{1,2}$-$M'_2$, as shown in Fig. 8, with $D'_{1,2}$ containing three **E**-dependencies, one for the system, two for the newly introduced functions in test modes. It is worth noting that in the transition of $D_{1,2}$ into $D'_{1,2}$, we do not revisit matrix elements that have been previously considered, but only those that are considered 'new' to $D_{1,2}$ (matrix cells are highlighted in blue), either due to the newly added entities (headers colored in red) or revised existing entities (not applicable in this rationalization).

Before moving on, it is worth revisiting the two resolutions to discuss what has happened technically during the synchronization. For either of the two approaches, essentially, what has been done can be understood as an interpretation of the $\mathbf{I}_\leftarrow$-dependencies into the semantics of the in-model dependencies in a UCD. In some cases, such interpretation becomes a direct mapping. For instance, in the first resolution, the $\mathbf{I}_\leftarrow$-dependencies are interpreted into Usage relations that can be modelled by «use». In other cases, such interpretation requires additional entities to be introduced before logical dependencies can be directly mapped into the semantics of in-model dependencies. For instance, in the second resolution, $\mathbf{I}_\leftarrow$-dependencies are interpreted into shared inclusions after adding the included Use Cases in both models.

Now, we move on to the rationalization of UCD-1 and UCD-3, beginning with the identification of logical dependencies to derive a $D_{1,3}$ from the triad, $M'_1$-$D^0_{1,3}$-$M^0_3$. Due to space limitation, we will not depict more matrices, but speak of specific matrix elements, using the convention, $\{M_i(e_x), M_j(e_y)\} = $ Dependency Type.

Here, it is self-explanatory that instead of using the initial UCD-1 as in Fig. 5 and represented in $M^0_1$, we will be using the revised UCD-1 in Fig. 7 and represent it by $M'_1$ to continue the thread of synchronization. Between $M'_1$ and $M^0_3$, one could argue that there is an $\mathbf{I}_\rightarrow$ dependency from 'Pick Part' (as well as 'Place Part') to 'Move Arm', i.e., in terms of corresponding matrix elements, we have

$$\{M'_1(U_1), M^0_3(U_1)\} = \mathbf{I}_\rightarrow;$$
$$\{M'_1(U_2), M^0_3(U_1)\} = \mathbf{I}_\rightarrow.$$

These reflect the concept of the Robotic Arm needing to move to the designated point before performing the pick and place function, and likely moving away from the point after the completion of the function.

Following the idea of interpreting the cross-model dependencies into the semantic of in-model dependencies discussed previously, we could interpret these $\mathbf{I}_\rightarrow$ dependencies into Usage dependencies. For instance, the completion of 'Pick Part' function would require the use of the 'Move Arm' function. As such, the resolution is modelled by repeating the 'Move Arm' Use Case in UCD-1 together with two «use» dependencies from 'Pick Part' and 'Place Part' Use Case to the 'Move Arm' Use Case. This is shown on the left in Fig. 9, which is rationalized with the original UCD-3, reproduced on the right of Fig. 9. Note that, the same resolution can also be represented by repeating the 'Pick Part' and 'Place Part' Use Cases in UCD-3 with the «use» dependencies, but this is graphically more complicated.

In this resolution, it is realized that in addition to an interpretation into a Usage dependency, there is also the precedence semantic, with 'Move Arm' likely performed both before and after 'Pick Part' (or 'Place Part) is performed. However, UCDs do not exhibit any explicit semantics for precedence. Hence, one cannot accurately and completely resolve these Implications into Equivalences without introducing new semantics for UCDs through extensions (not the preferred method in general), or using other types of diagrams, such as Activity and Sequence diagrams, which focus on precedence (the preferred method). Since we have scoped the case study to UCDs, we will be neglecting the precedence part of the interpretation to move forward with the rationalized UCDs shown in Fig. 9. As such, $D_{1,3}'$ in the synchronized $M_1'$-$D_{1,3}'$-$M_3'$ triad (with $M_3' = M_3^0$, since no change has been made to UCD-3) would contain two **E**-dependencies:

$$\{M_1'(\mathbb{S}), M_3'(\mathbb{S})\} = \mathbf{E};$$
$$\{M_1'(U_7), M_3'(U_1)\} = \mathbf{E}.$$

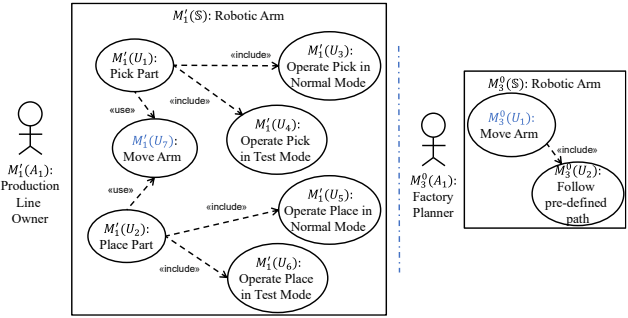

**Fig. 9.** Rationalized UCD-1 and UCD-3.

Finally, we will be rationalizing the last pair, UCD-2 and UCD-3. Following the same processes demonstrated previously, we arrive at a pair of rationalized UCDs depicted in Fig. 10 and with $D_{2,3}'$ in the synchronized $M_2'$-$D_{2,3}'$-$M_3'$ triad containing two **E**-dependencies:

$$\{M_2'(\mathbb{S}), M_3'(\mathbb{S})\} = \mathbf{E};$$
$$\{M_2'(U_4), M_3'(U_4)\} = \mathbf{E}.$$

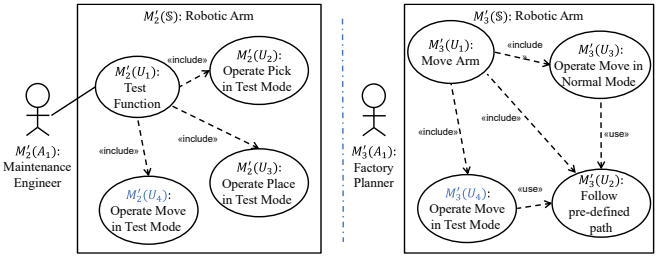

**Fig. 10.** Rationalized UCD-2 and UCD-3.

In this step of rationalization, one would raise a concern on how the changes made to UCD-2 and UCD-3 might affect the rationalized UCD-1. This is where one can trace trough the synchronized triads developed so far to identify where iteration is needed. For instance, the introduction of $M_2'(U_4)$ in $M_2'$ would impact the $M_1'$-$D_{1,2}'$-$M_2'$ triad by introducing a new column (with $M_2'(U_4)$ as the header) in the $D_{1,2}'$ matrix. One would then need to go through the new cross-model entity pairings (c.f., Fig. 8), e.g., $\{M_1'(U_1), M_2'(U_4)\}$ to see if there are new cross-model dependencies that could arise. In this case study, the rationalized UCD-2 and UCD-3 as in Fig. 9 do not introduce any new cross-model dependencies with respect to UCD-1 that requires further synchronization. Therefore, we claim that all three UCDs are now rationalized with a set of three synchronized triads: $M_1'$-$D_{1,2}'$-$M_2'$, $M_1'$-$D_{1,3}'$-$M_3'$, and $M_2'$-$D_{2,3}'$-$M_3'$. Due to space limitations, final matrices are not depicted.

In this case study, the order for which pair-wise rationalization was conducted followed a natural order with ascending index. In principle, the order followed should not make a difference to the results. However, certain orders are likely to be less efficient than others if the synchronization of models in a later step introduces changes that desynchronize the models that were synchronized in previous steps. Again, in the case study, such desynchronization did not occur. Therefore, designing an efficient order based on available information would be valuable for accelerating the overall rationalization process, as if we are designing a 'system of stakeholder requirements'.

*C. Seamless Synthesis by Merge*

With the set of synchronized triads, we can now perform the synthesis process by a straightforward merge based on the framework for synthesis introduced in Section II-B. In simple terms, equivalent Use Cases appearing in multiple diagrams are merged into one Use Case and the merged Use Case inherits all the in-model dependencies that the pre-merge Use Cases possess. For instance, the 'Move Arm' Use Case appeared in both rationalized UCD-1 and UCD-3, with IDs $M_1'(U_7)$ and $M_3'(U_1)$ respectively. After the merge, the two Use Cases are merged into one, as depicted in the synthesized UCD, ***M***, in Fig. 11, inheriting two «use» dependencies from UCD-1 ($M_1'$) and three «include» dependencies from UCD-3 ($M_3'$). To maintain traceability, merged model entities have multiple IDs as inherited from the siloed UCDs.

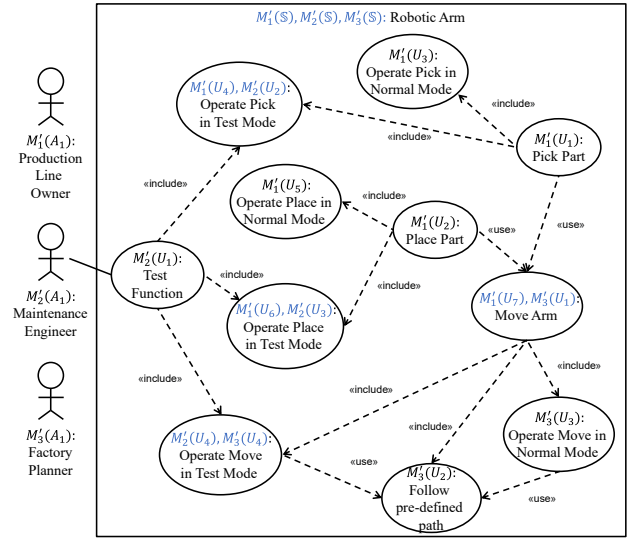

**Fig. 11.** The synthesized UCD.

IV. CASE STUDY PART II – HANDLING CHANGES

In the case study, we have demonstrated how synchronization enables rationalization. Any change from any stakeholder, indexed by $i$, would then be reflected as a change to the corresponding UCD and its matrix representation, $M'_i$. Such a change could desynchronize the models, similar to the way in which changes introduced during a rationalization step could desynchronize previously rationalized models. Hence, any changes made after the rationalization and synthesis would require agreement among stakeholders, in terms of re-rationalizing the models. The scale of re-rationalization required depends on the nature and scale of the change. Re-rationalization, at whatever scale, could be achieved by the synchronization concept using matrix triads, as described in Section III. In brief, while making changes to the previously synchronized triads, stakeholders are prompted to identify new cross-model dependencies with respect to the change made. In this section, we illustrate how this synchronization of changes is practically achieved, building on the case study used so far.

*A. A New Stakeholder*

The first change we introduce to the case study is a new stakeholder group that was intentionally left out in the initial elicitation. This stakeholder group consists of a Safety Analyst and the Part Inspector, sharing a goal that the Robotic Arm will operate safely in a collaborative manner with the Part Inspector. Specifically, the Robotic Arm would place a part on the workbench for the Part Inspector to inspect and pick the part up when it passes human inspection.

Safety was previously considered by the Factory Planner, who considered that the Robotic Arm needs to follow pre-defined paths so that, in principle, it will always be at a safe distance from any workers involved in production. However, a safety analyst would notice that even with such intent, there is still a safety hazard: namely when the Part Inspector is located too close to a Robotic Arm, possibly due to making a human error. Eventually this hazard could lead to an accident where the Robotic Arm harms the worker. To address this safety concern, the need from this stakeholder group could be elicited into a UCD, denoted as UCD-4 and depicted in Fig. 12. In brief, the Robotic Arm needs to stop any movement when any workers are within close and unsafe proximity.

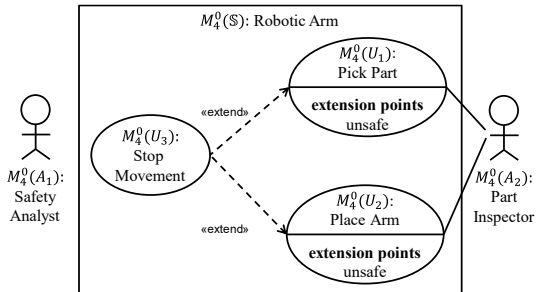

**Fig. 12.** UCD-4 for the new stakeholder group: safety analyst and factory workers.

The rationalization of UCD-4 with the other UCDs follows a similar rationalization process to that introduced in Section III, with the difference being that the previous UCDs are already rationalized. Instead of going through the details, we highlight the key outcomes below.

Firstly, with respect to $M'_1$, identifying cross-model dependencies in $D_{1,4}$ would yield the following matrix elements:

$$\{M_4^0(\mathbb{S}), M'_1(\mathbb{S})\} = \mathbf{E};$$
$$\{M_4^0(U_1), M'_1(U_1)\} = \mathbf{E};$$
$$\{M_4^0(U_2), M'_1(U_2)\} = \mathbf{E};$$
$$\{M_4^0(U_3), M'_1(U_7)\} = \mathbf{C}.$$

The first three **E**-dependencies are self-evident due to the repetition of Use Case naming convention. Strictly speaking, the 'Pick Part' and 'Place Part' Use Cases are modified with adding an extension point, which needs to be repeated as well in $M'_1$, if **E**-dependency is confirmed and agreed between the stakeholders. The last dependency in the list is a contradiction, **C**-dependency, because the Robotic Arm cannot move and stop moving at the same time.

To resolve this contradiction, one would need to introduce a prioritization on which function is to be performed under what conditions. In principle, safety should be prioritized in all circumstances. One could reflect this prioritization by including an extension point for the 'Move Arm' Use Case, such that it will also trigger 'Stop Movement' when the situation is considered unsafe. However, a proper resolution would again require the usage of other diagram types, such as Activity diagrams to allow the priority logic to be properly modelled, e.g., by using Disruption nodes and Merge nodes.

Next, with respect to $M'_2$, for $D_{2,4}$, we have a similar $\mathbf{I}_{\leftarrow}$-dependency concerning the 'Testing Function' Use Case. This can be resolved again by introducing an 'Operate Stop in Test Mode' and 'Operate Stop in Normal Mode' pair that would not desynchronize any other models (c.f., Fig. 7 and Fig. 9).

Finally, with respect to $M'_3$, we would have the contradiction issue with the 'Move Arm' Use Case, but this has been already resolved during the synchronization with $M'_1$.

With the above treatment, UCD-4 would be rationalized with the other UCDs and a new $\boldsymbol{M}$ could be seamlessly synthesized again by merging $M'_4$ along with the set of $D$-matrices into the previous $\boldsymbol{M}$. It is worth noting that the treatment of UCD-4 in this section suggests another approach to the ordering and iterative synchronization of the models. Instead of synchronizing one model with the rest of the models and then iterating from the next model in the list (as in Section III-C), this other approach can be described as iteratively adding a new model and synchronizing it with the existing synchronized models.

*B. Contradicting Needs*

The resolution of the contradicting needs adopted in the previous subsection assumed prioritizing safety over production throughput. However, the Factory Planner may argue that such a prioritization could potentially lead to Robotic Arms stopping too often, resulting in the production throughput not generating a minimum income to sustain the production line. Therefore, such a resolution cannot be accepted during the synchronization of $M'_3$ and $M_4^0$. This is an example for which the resolution of a conflicting need between two stakeholders creates a new conflicting need with the third stakeholder.

Our proposed rationalization framework and the synchronization process facilitates the identification of such situations. Specifically, as described in the previous subsection, the resolution following prioritization of safety results in the 'Stop Movement' Use Case to be added to UCD-1, which will be extending the 'Move Arm' Use Case, $M'_1(U_7)$. This added dependency flags up a change of property for the 'Move Arm' Use Case, as such, the matrix element, $\{M'_1(U_7), M'_3(U_1)\} = \mathbf{E}$, in $D'_{1,3}$, requires a re-examination of the validity of the $\mathbf{E}$-dependency. As discussed earlier, this dependency no longer holds due to the concern of financial feasibility from the Factory Planner.

An appropriate resolution would then require stakeholder consultation involving all parties. To avoid repeating the issue where resolution of conflicts generates new conflicts, a possible resolution here is to continue prioritizing safety, but requesting the Robotic Arm to redirect itself to avoid unsafe situations. This would then require Artificial Intelligence, which could be modelled by changing the 'Follow Pre-defined Path' Use Case, $M'_3(U_2)$, into a 'Follow Self-defined Path' Use Case. Again, this change, like any other, needs to be investigated through the synchronized triads to determine whether they desynchronize the models by creating new cross-model dependencies other than the $\mathbf{E}$-dependency. Through a synchronization enabled change management process, stakeholders will eventually arrive at an agreement with reasonable confidence reflected by the achieved synchronization.

**Table I Summary of Case Study Results**

| Model Pairs | Identified Incompatibility | Type | Proposed Resolution |
|---|---|---|---|
| M1&M2 | Unclear provision of testing function for 'Pick Part' functionality | **I**-Dep. | Introduce normal mode and test mode for 'Pick Part' functionality |
| M1&M2 | Unclear provision of testing function for 'Place Part' functionality | **I**-Dep. | Introduce normal mode and test mode for 'Place Part' functionality |
| M1&M3 | Missing dependencies | **I**-Dep. | Model the missing dependencies using «use» |
| M2&M3 | Unclear provision of testing function for 'Move Arm' functionality | **I**-Dep. | Introduce normal mode and test mode for 'Move Arm' |
| M1&M4 | Contradicting command on arm movement | **C**-Dep. | Prioritise safety requirement |
| M3&M4 | Potentially infeasible to prioritise safety | **C**-Dep. | Redefine 'Move Arm' lower-level details |

To summarize the case study as presented in Section III and IV, incompatible needs that were identified during rationalization are briefly captured in Table I, along with the type of incompatibility and proposed resolution. In the initial rationalization of three UCDs, no contradiction was identified, but there were four structural incompatible cases due to missing relationships between Use Cases derived from different stakeholders. Following the change introduced in the 4th UCD, a contradiction has been identified. Initial resolution leads to a further contradiction that can be resolved by pushing the changes into further design details. In this simple case study, it is difficult to argue that these incompatible cases are only discoverable through our proposed processes. However, the frameworks offer a systematic and rigorous approach that promises a level of automation, e.g., tracking of effects from changes brought by proposed resolutions into other models, to avoid human errors.

## V. APPLICABILITY TO SELECTED WORKS

In this section, we discuss related work and comment on the applicability of the framework and the process to these works. This will also serve to further validate our model theoretic framework.

### A. Different Diagram Types

There are numerous existing works in model-based requirements engineering that offer a variety of methods for eliciting stakeholder needs using standardized modelling languages. The development of UCD-based approaches can be traced back to the early 90's [28], when object-oriented software engineering gained its growing popularity. Since then, several applications using approaches based on UCDs have been reported in various domains, e.g., Data Warehouse Systems [29] and the IoT [30]. From the UCD-based case study, it is clear that our approach naturally aligns with these kinds of approaches.

It has been acknowledged by the community that although UCDs are a good starting point for needs elicitation, they are not always the best diagram type to work with due to lack of semantics such as precedence. For instance, early work by Waite and Logan [31], utilized ADs for user need elicitation, which captured clearly intended functional needs as well as flows and physical objects that need to be generated in the process with an intuitive restaurant servicing example. Following this line of thought, to cope with the limitations of a single diagram type, various types of diagrams are typically used in conjunction with each other to enable full scope elicitation [32-35]. Multi-diagram types become even more demanding when concerning System of Systems due to complex interactions and the need for taking the logical architecture of existing constituent system into account in the elicitation process [36, 37]. While our framework can handle different diagrams, since logical dependencies used to synchronize models are diagram-independent, when multiple diagrams of different types are used in combination, this creates a situation that requires 'two-dimensional synchronization' (Fig. 13).

Interestingly, a number of the related works emphasized the importance of a model-based approach being able to manage and resolve conflicts in addition to just facilitating stakeholder communications [33, 36], resonating with what we proposed in this work.

A particular work of interest is the one by Kaiya, Osada and Kaijiri which utilizes derived UCDs as the basis for non-functional requirements elicitations [39]. A key step in their approach is to 'find common and/or similar use cases in the UCDs' and 'find differences among the surroundings', which essentially is what we are trying to achieve: rationalization.

*B. Two-dimensional Synchronization*

The discussion in the previous subsection leads to a potentially demanding capability for model-based requirements elicitation to support the wider practice of MBSE: two-dimensional synchronization, which is an idea describing simultaneous 'horizontal' and 'vertical' synchronization of models. Here, horizontal synchronization refers to the process of endogenous rationalization as described in this paper, while vertical synchronization refers to exogenous elaboration that was explored in a previous paper on model synchronization [4].

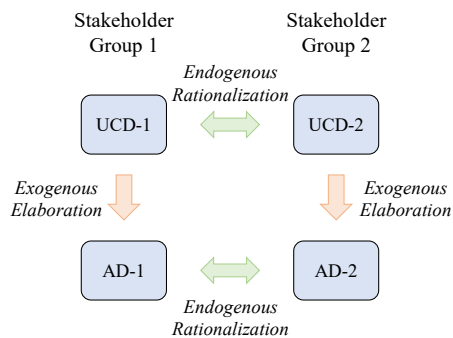

**Fig. 13.** Illustration of two-dimensional synchronization.

This idea of two-dimensional synchronization is conceptually depicted in Fig. 13, using UCDs and Activity Diagrams (ADs) for illustration. As shown, let us consider two stakeholder groups, denoted as 1 & 2 with the stakeholder needs elicited in a siloed manner, but now with multiple diagram types, UCD and AD in this illustration, to fully articulate their intent. Following the framework introduced in Ref. [4], the AD can be developed using structure preserving transformations. The resultant AD is eventually an elaboration of the UCD in a consistent way such that the UCD and the AD form a synchronized pair, e.g., UCD-1 and AD-1. Now that the needs for each stakeholder group are modelled in a UCD and AD pair, the new rationalization process would require synchronization between the UCDs and between the ADs. The new process is intuitive but will be complicated because while synchronizing horizontally to achieve rationalization, the vertical synchronization between UCD and AD pairs needs to be maintained. We envision that a combined framework based on Ref. [4] and this work would help understanding of the complexity, but the detailed executable process requires further research to ensure applicability. The complexity would need to be handled with a sufficient level of automation, if such a process were to be adopted in practice.

## VI. Conclusion

Intended outcomes of stakeholder needs modelling and analysis to enable system requirements definition include resolution of requirements conflicts to achieve agreement between stakeholders. We have a specified model theoretic framework that facilitates rationalization and synthesis of outputs from stakeholder needs elicitation. The framework has been demonstrated through a case study on robotic arms in which siloed stakeholder needs are synthesized and further re-rationalized based on stakeholder changes. The demonstration exhibits both an efficacy and robustness of our matrix-based rationalization and seamless synthesis process.

The real test of the efficacy and robustness though will be in the application of the framework and matrix-based processes to complex systems and systems of scale. As demonstrated in this paper, seamless synthesis was for models that are of the same type of semantic structure. It is natural that different stakeholder groups might use different types of models and diagrams or modelling languages, including domain-specific languages rather than general purpose languages. Further research is then needed on how exogenous synthesis can be performed across rationalized models of different types.

Future work is planned to address both research and commercialization. The advanced mathematics underlying the model theoretic framework will be published and provide a basis for collaboration with commercialization partners such as the Object Management Group. Longer term collaboration efforts will be focused on elaboration and implementation of the framework for MBSE practices in aerospace systems.

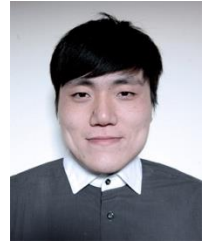
**SIYUAN JI** (M'16) received the PhD in physics from University of Nottingham, Nottingham, UK in 2015. He previously received the MSc degree in physics from the same institution.

He is a Senior Lecturer in Systems Engineering at Loughborough University. Previously, he was a Lecturer and the Programme Lead for MSc in Safety-Critical Systems Engineering at the Department of Computer Science at the University of York, UK. His research is focused on model-based systems engineering and system safety assessments, and constraint driven design algorithms.

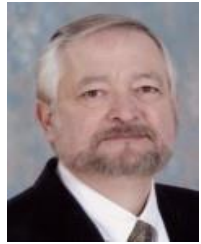
**CHARLES E. DICKERSON** (M'06–SM'17) received the PhD in mathematics from Purdue University, West Lafayette, Indiana, USA in 1980.

He is Professor and Chair of Systems Engineering at Loughborough University, UK. Previously he was Technical Fellow in BAE Systems in the USA. His aerospace experience further includes the Lockheed Skunkworks and Northrop Advanced Systems. He was also a member of the Research Staff at Lincoln Laboratory, Massachusetts Institute of Technology. He has served secondments as the Aegis Systems Engineer for US Navy Ballistic Missile Defense; and Director of Architecture for the Navy's Chief Engineer. He is currently Co-chair of Mathematical Frameworks in the Object Management Group (OMG).

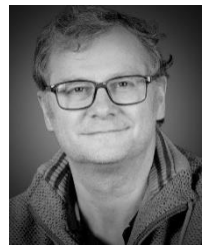
**MICHAEL WILKINSON** received the PhD in theoretical physics from King's College, London, UK in 1985. Previously, he received the BSc in physics from the same institution.

He is Chief Technologist and Chief Systems Engineer at BAE Systems Maritime, UK, and is a visiting Professor at Loughborough University. Previously he was a Technical Director and Professional Head of Discipline for Systems Engineering at Atkins. Prior to that he was Technical Director of the Niteworks partnership of the MOD. He has served as President and Academic Director of the UK Chapter of the International Council on Systems Engineering. He is currently Co-Chair of the INCOSE UK Chapter Architecture Working Group.